# Improved Fair-Zone technique using Mobility Prediction in WSN

K.Ramesh[1] and Dr. K.Somasundaram[2]

[1]Dept of ECE, Nandha Engineering College, Erode.

rameshk.me@gmail.com

[2]Dept of CSE, Jaya Engineering College, Chennai.

soms72@yahoo.com

## ABSTRACT

*The self-organizational ability of ad-hoc Wireless Sensor Networks (WSNs) has led them to be the most popular choice in ubiquitous computing. Clustering sensor nodes organizing them hierarchically have proven to be an effective method to provide better data aggregation and scalability for the sensor network while conserving limited energy. It has some limitation in energy and mobility of nodes. In this paper we propose a mobility prediction technique which tries overcoming above mentioned problems and improves the life time of the network. The technique used here is Exponential Moving Average for online updates of nodal contact probability in cluster based network.*

## 1. INTRODUCTION

Wireless Sensor Network (WSN) is the fast grooving wireless technique which consists of number of sensor nodes and a sink (fixed base station). Sensor nodes are capable of sensing or measuring the physical data of the area to be monitored. Then the nodes will aggregate the sensed data, store and forward it to the sink through air interface. Nodes may be scattered to monitoring area or may be mobile terminal or combination of the two.

Clustering is especially important for sensor network applications where a large number of ad-hoc sensors are deployed for sensing purposes. If each and every sensor starts to communicate and engage in data transmission in the network, great network congestion and data collisions will be experienced. This will result to drain limited energy from the network. Node clustering will address these issues. Scalability of the network of those WSN is useful to meet load balancing and efficient resource utilization constraints. In cluster networks, sensors are partitioned into smaller clusters and cluster head (CH) for each cluster is elected. Sensor nodes in each cluster transmit their data to the respective CH and CH aggregates data and forward them to a central base station. Clustering through creating a hierarchical WSN facilitates efficient utilization of limited energy of sensor nodes and hence extends network lifetime. Although sensor nodes in clusters transmit messages over a short distance (within clusters), more energy is drained from CHs due to message transmission over long distances (CHs to the base Station) compared to other sensor nodes in the cluster. Periodic re-election of CHs within clusters based on their residual energy is a possible solution to balance the power consumption of each cluster. In addition, clustering increases the efficiency of data transmission by reducing the number of sensors attempting to transmit data in the WSN, aggregating data at CHs via intra-cluster communication and reducing total data packet loses.

Low-Energy Adaptive Clustering Hierarchy (LEACH) [2] is the most popular cluster-based routing protocols in Wireless Sensor Networks. In LEACH the cluster heads are randomly selected and when the cluster head die then another node will be selected as cluster head. Therefore, the cluster head role is kept on rotating to balance the energy dissipation of the sensor nodes in the networks. As a result, the nodes cannot provide end to end delivery of the information. At the same time nodes break up their communication if they move out of the coverage area.

## 2. CHALLENGING ISSUES IN CLUSTERING ALGORITHM

Following are the some important challenging [1] issues in cluster based WSNs:

• *Limited Energy:* Wireless sensor nodes have limited energy storage and once they are deployed, it is not practical to recharge or replace their batteries. With the capability of reducing the amount of data transmission, the clustering algorithms are more energy efficient compared to the direct routing algorithms. This can be achieved by balancing the energy consumption in sensor nodes by optimizing the cluster formation, periodically re-electing CHs based on their residual energy, and efficient intra-cluster and inter-cluster communication. But clustering algorithms should prevent high energy cluster reconstruction process.

• *Network Lifetime:* The energy limitation on nodes results in a limited network lifetime for nodes in a network. Clustering technique helps to prolong the network lifetime of WSNs through reducing the number of nodes contending for channel access, data aggregation at CHs via intra-cluster communication and direct or multi-hop communication by CHs with a base station. Proper design should focus on increasing network lifetime.

•*Cluster formation and CH selection:* Cluster formation and CHs selection are two of the important operations in clustering algorithms. Energy wastage in sensors in WSN due to direct transmission between sensors and a base station can be avoided by clustering the WSN. Clustering further enhances scalability of WSN in real world applications. Selecting optimum cluster size, election and re-election of CHs, and cluster maintenance are the main issues to be addressed in designing of clustering algorithms. The selection criteria to isolate clusters and to choose the CHs should maximize energy utilization, as well as function for a variety of applications.

• *Synchronization:* Slotted transmission schemes like TDMA allow nodes to regularly schedule sleep intervals to minimize the energy used. Such schemes require synchronization mechanisms to setup and maintain the transmission schedule and the effectiveness of this mechanism must be considered.

• *Data Aggregation:* Data aggregation allows the differentiation between sensed data and useful data. In a densely populated network there are often multiple nodes sensing similar information

• *Node Mobility:* Due to the nature of Wireless Sensor Networks, they are often prone to node mobility, node death and interference. All of these situations can result in link failure. When looking at clustering schemes, it is important to look at the mechanisms in place for link recovery and reliable data communication. Mobile Cluster Head (CH) node in the network leads to big issue. Proposed method tries to predict the mobility of nodes.

## 3. OPTIMIZED FZ – LEACH (OFZ-LEACH)

One of the issue in basic LEACH [2][3] is the hidden cluster. Hidden cluster is the areas were no cluster member is eligible to act as a Cluster Head due to low average power. When hidden clustering happened, information on those nodes unable send sensed data to base. They are simply like a dead node. Fair zone is concept introduced in [4] overcome this issue by selecting a CH outside the cluster. FZ- LEACH [4] failed at node mobility issue and results in link failure.

In proposed method a mobility prediction technique is used to identify the Cluster Head movement. Nodal Contact Probability updated in online by sending message between nearby nodes and to apply Exponential Moving Average (EMA) for quick prediction of the mobile nodes. This can be used to detect the CH movement outside the coverage area. If energy level of CH is below the critical limit, then it is need to re-elect the CH and avoid the link failure. This improves the life time of the network. The algorithm is event-driven, where the key part lies on the meeting event between any pair of nodes. The set of functions in the algorithm including Sync, Leave and Join is outlined below. The Methodology will give an idea about how the algorithm performs its function.

• *Methodology:*

1. The base station (i.e. sink node) is located inside the sensing field.
2. Nodes are location-unaware, i.e. not equipped with GPS capable antennae.
3. Communication within the square area is not subjected to multipath fading.
4. The communication channel is symmetric.
5. Data gathered can be aggregated into single packet by Cluster Heads (CH).
6. Nodes are left unattended after deployment. Therefore battery re-charge is not possible.
7. An Exponentially Weighted Moving Average (EWMA) scheme is employed for on-line updating nodal contact probability.
8. Weighting factors which decrease exponentially. The weighting for each older data point decreases exponentially, giving much more importance to recent observations while still not discarding older observations entirely.
9. True contact probability. Subsequently, a set of functions including *Sync()*, *Leave()*, and *Join()* are devised to form clusters and select gateway nodes based on nodal contact probabilities.
10. The cluster table consists of four fields: Node ID, Contact Probability, Cluster ID, and Time Stamp.
11. Each entry in the table is inserted/ updated upon meeting with another node, by using the aforementioned online updating scheme.
12. The gateway table, used for routing, consists of four fields: Cluster ID, Gateway, Contact Probability, and Time Stamp.

• *Nodal Delivery Probability:*

The delivery probability indicates the likelihood that can deliver messages to the sink. The delivery probability of a power *i,* is updated as follows,

$$\xi_i = \begin{cases} (1-\alpha)[\xi_i] + \alpha\xi_k & \text{Transmission} \\ (1-\alpha)[\xi_i] & \text{Timeout} \end{cases} \quad (1)$$

where $\xi_i$ is the delivery probability of power *i* before it is updated, $\xi_k$ is the delivery probability of node *k* (a neighbour of node *i*), and α is a constant employed to keep partial memory of historic status.

*Sync()*

The *Sync()* process is invoked when two cluster members meet and both pass the membership check. It is designed to exchange and synchronize two local tables. The synchronization process is necessary because each node separately learns network parameters, which may differ from nodes to nodes. The Time Stamp field is used for the "better" knowledge of the network to deal with any conflict.

*Leave()*

The node with lower stability must leave the cluster. The stability of a node is defined to be its minimum contact probability with cluster members. It indicates the likelihood that the node will be excluded from the cluster due to low contact probability. The leaving node then empties its gateway table and reset its Cluster ID.

*Join ()*

The *Join ()* procedure is employed for a node to join a "better" cluster or to merge two separate clusters. A node will join the other's cluster if

1. It passes membership check of all current members.
2. Its stability is going to be improved with the new cluster. By joining new cluster, it will copy the gateway table from the other node and update its cluster ID accordingly.
3. Thus the distributed clustering algorithm is used to form a cluster in DTMN (Disruption-Tolerant Mobile Networks).

## 4. EXPERIMENTAL SETUP

The performance of power balanced communication scheme is evaluated using Network Simulator-2, which simulate node mobility, realistic physical layer radio network interface and AODV protocol. Two Evaluation methods proceeded. First one is based time Vs other parameters which is on the simulation of 50 nodes located in the area of 1500 x 300 m^2. Second one is based on Number of Nodes used Vs other parameters. The traffic simulator is Constant Bit Rate (CBR). The three different schemes used for comparison are, LEACH [2] (equal cluster size) FZ – LEACH [4] (non EMA) and OFZ LEACH (power balanced communication with EMA).

The performance metric used in this research are throughput, packet delivery ratio, bandwidth end to end delay, energy construction are based on number of node analysis and Average Number of

International Journal Of Advanced Smart Sensor Network Systems ( IJASSN ), Vol 2, No.2, April 2012

Cluster Member and Cluster Head are based on time. The various measures and details of the all parameters are given below.

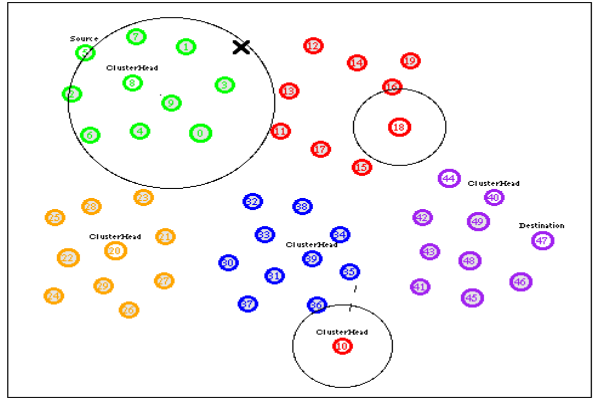

Figure 1. Snapshot for Fair-Zone

### A.     Average Number of Cluster Member:

Average Number of Cluster Member is decided by number of clusters and members in the clusters. Low average values leads by large amount of variation in cluster size. In Figure 2 proposed method has very low average as compared with others. It intimates that very low and very high size clusters are distributed in the proposed method.

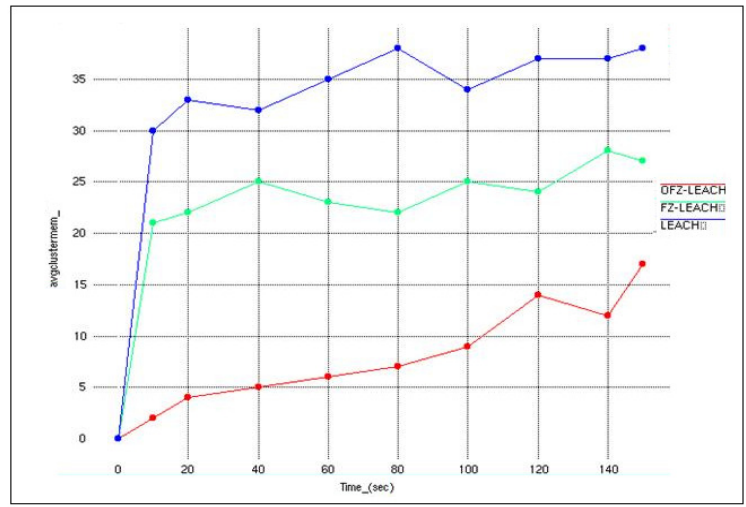

Figure 2. Performance comparison using Average Number of Cluster Member

### B. Average Number of Cluster Head:

Average Number of Cluster Head decides the life time of the system. Figure 3 shows the comparison of Average Number of Cluster Head for specific duration. It shows that normal LEACH has high average of heads from the starting. It leads to the low life time of that system. As compared with FZ – LEACH, the proposed method has low average value and it increases life time of new system.

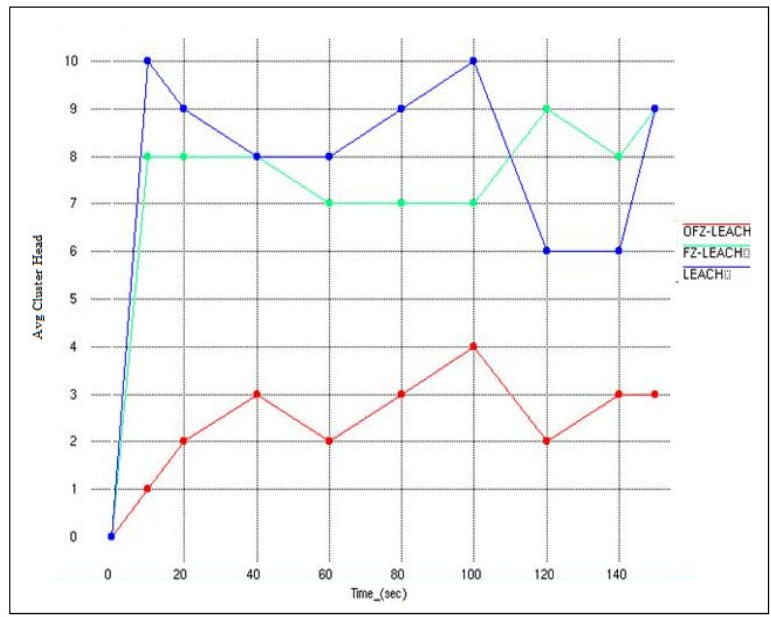

Figure 3. Performance comparison using Average Number of Cluster Header

### C. Packet Delivery Ratio:

It is defined as the percentage of the ratio of number of packets received to the number of packets sent.

$$PDR = \frac{Number\ of\ packets\ received}{Number\ of\ packet\ sent} \times 100\ \% \qquad (2)$$

The graph shows in Figure 4 is the variation of Packets (in Kbytes) received based on time when two different routing schemes are implemented.

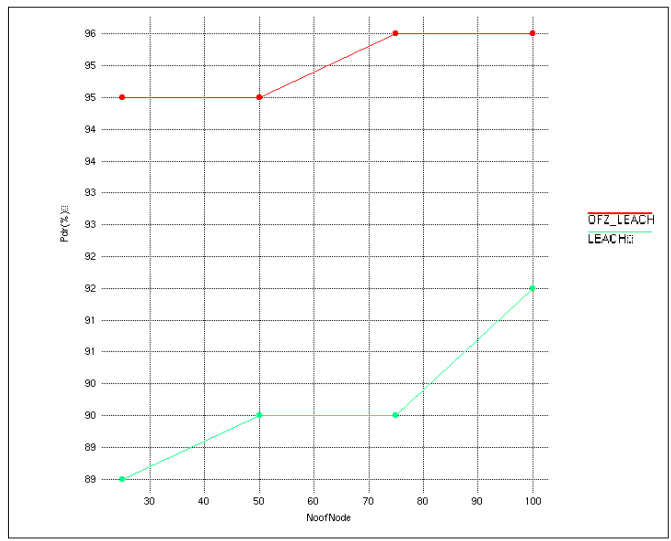

Figure 4. Performance comparison using Packet delivery ratio.

## D.     End to End Delay:

The time interval between the first packet and second packet is called End to End Delay or latency. Delay is the most important parameter for sensitive application and here the proposed method has very low delay and other two methods are comparatively high delay. It is shown in Figure 5.

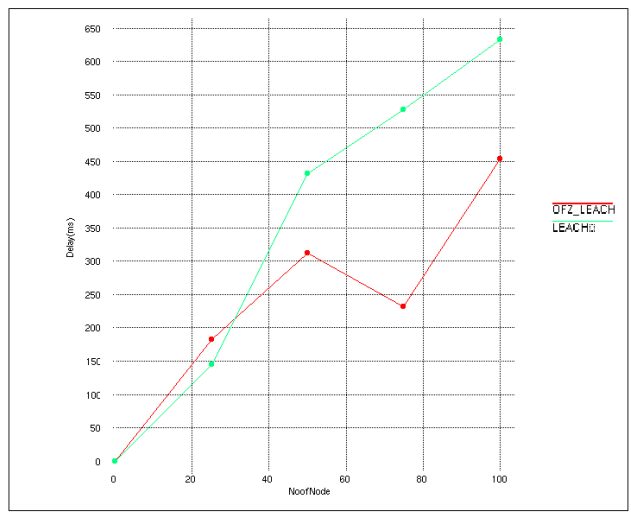

Figure 5. Performance comparisons using end to end latency
.

*E.     Throughput:*

Throughput is the ratio of number of packets received to the specific time period. Throughput needs to be high for a better system. The proposed OFZ – LEACH was comparatively high throughput as shown in Figure 6.

$$Throughput = \frac{Number\ of\ packets\ received}{Time(sec)} \times 100\% \qquad (3)$$

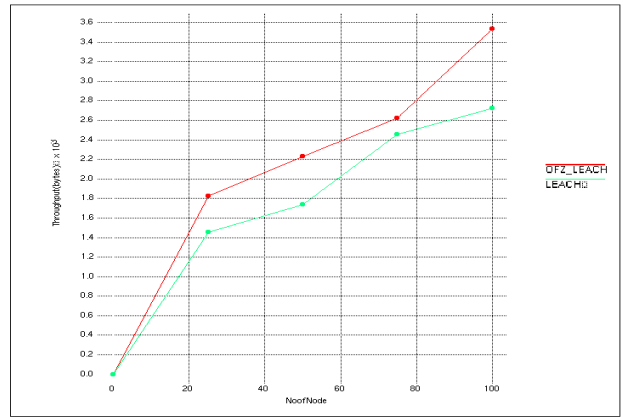

Figure 6. Performance comparison using Throughput

*F.     Energy- Consumption:*

Average residual energy of the nodes is another impartment parameter to increase the life time of WSN. In the proposed method residual energy of nodes is increased from the previous methods like LEACH and FZ – LEACH. Figure 7 shows the energy consumption for different number of node combinations.

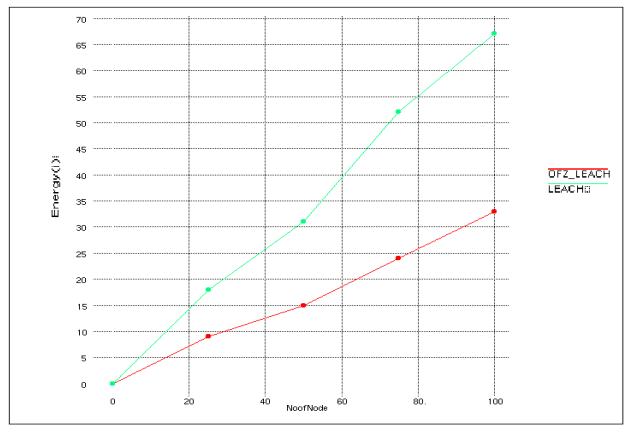

Figure 7. Performance comparison using Energy- Consumption

## 5. CONCLUSIONS

Establishing end-to-end connections for data delivery among Wireless Sensor Networks becomes impossible as communication links only exist temporarily. In such networks, routing is largely based on nodal contact probabilities. To solve this problem, an exponentially moving average (EMA) scheme is employed for on-line updating nodal contact probability. In proposed OFZ-LEACH algorithm, which is based on the original LEACH protocol and considers Far-Zone concept. The results proves the improvement in the performance in the original LEACH protocol in terms of energy dissipation rate and network lifetime. A set of functions including sync(), leave(), and join() are devised for cluster formation and gateway selection. Finally the gateway nodes exchange network information and perform routing. The results have shown that it achieves higher delivery ratio and significantly lower overhead and end-to-end delay, compared with its non-EMA. Finally proposed method try to reduce the end to end delay and increase the life time of the system.

## Authors

**Prof. K.Ramesh** acquired B.E. degree in ECE from Bharathiar University in 2000. He obtained M.E. in Computer Science and Engineering from Anna University in 2009. He is having teaching experience about 10 years. He is currently serving as Associate Professor in Electronics and Communication Engineering department at Nandha Engineering College. He published 2 papers in International journals and presented 10 papers in refereed National & International Conferences. His research interests include networking, security and wireless sensor networks.

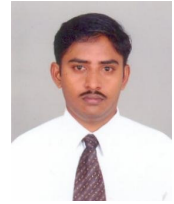

**Dr.K.Somasundaram** acquired B.E. degree in ECE from Mananonmanium Sundaranar University in 1994. He obtained M.E. in Computer Science and Engineering from Madurai Kamaraj University in 1998 and completed PhD in Computer science and Engineering from Anna University, Chennai in 2009. He is having industry and teaching experience about 17 years. He served various positions in industry and Teching. He is currently serving as Professor in Computer Science and Engineering department at Aarupadai Veedu Institute of Technology, Chennai. He published 10 papers in International journals and presented 30 papers in refereed national & International Conferences. He is guiding 4 PhD scholars and guided more than 23 M.E., Thesis. He is a member of IEI, CSI, ISTE, C.Engg(IEI).

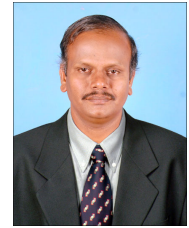